%% file: ZadoffChu-Tutorial_v5.tex
\documentclass[12pt, draftclsnofoot, onecolumn]{IEEEtran}
\usepackage{amsmath}
\usepackage{amssymb}
\usepackage{graphicx}
\usepackage{color}
\usepackage[latin1]{inputenc}

\include{JeffNotation}

\def\Nzc{\mathcal{N}_{\rm zc}}
\def\Nsf{\mathcal{N}_{\rm sf}}

\begin{document}

\title{A Primer on Zadoff-Chu Sequences}
\author{Jeffrey G. Andrews
\thanks{J. G. Andrews (jandrews@ece.utexas.edu) is the Truchard Family Endowed Chair at the University of Texas at Austin. 
Date last revised: \today}}

\maketitle

\vspace{-0.6in} 
\section{Introduction and Overview}

Zadoff-Chu (ZC) sequences are an important manifestation of spread spectrum in modern cellular systems, including LTE and 5G NR.  They have to some extent displaced Pseudorandom Noise (PN) and Walsh sequences which were the mainstays of 3G cellular (WCDMA and cdma2000) and the 2G-era IS-95.    ZC sequences are complex sequences with unit amplitude and particular phase shifts, as opposed to Walsh and PN codes which are real and binary valued, most commonly $\pm1$.  In this primer, we define ZC sequences and introduce their key properties, provide some examples, and overview their uses in LTE and 5G.

ZC sequences have a number of remarkable and desirable properties that we define in the next section.  Because of these properties, they are used for many key functions in 4G LTE and 5G NR cellular systems, and are likely to be prevalent in future cellular systems as well.  In LTE and 5G, they are widely used for a number of important initial access and overhead channel functions that are often overlooked by engineers who focus on maximizing the data transmission rate and spectral efficiency.    For example, ZC sequences\footnote{Technically, they are extended ZC sequences, which are not strictly speaking, ZC sequences.  We explain this important subtlety in Sect. \ref{sec:ZC-like}.} are used for initial access in both the downlink (synchronization sequences) and uplink (random access preambles).   They are also used for transmitting uplink control information, and as pilot symbols for both uplink channel sounding and fine-grained channel estimation.    It is not an exaggeration to say that most types of signals other than the data transmissions in modern cellular standards utilize ZC sequences.

\section{Definition and Key Properties}

\subsection{Definitions}

\textbf{Zadoff-Chu sequence.}  A Zadoff-Chu sequence has two key parameters, (i) the root index $q = 1, 2, \ldots, \Nzc-1$ and (ii) the length of the sequence, $\Nzc$, which must be an odd number, and is often a prime number.   Given these two parameters, the $q^{th}$ ZC sequence $s_q[n]$ is defined as 
\begin{equation}
\label{eq:defn}
s_q[n]  = \exp\left[ -j\pi q \frac{n(n+1)}{\Nzc} \right]
\end{equation}
where $n  = 0, 1,2, \ldots,\Nzc - 1$.    Note that each sequence has length $\Nzc$, while the number of such sequences is $\Nzc-1$.    For example, if the ZC is used as a time domain spreading code as in Code Division Multiple Access (CDMA), the spreading factor would be $\Nsf = \Nzc$ and $n$ denotes discrete time, and each user utilizes a different value of $q$.

\textbf{Cyclic shifts and correlations.}   A \emph{cyclic shift}, also known as a circular shift, is simply the rotation of a finite length sequence.   Specifically, given a sequence $x[n]$ of length $N$, we shall write the $m$th cyclic shift of $x[n]$ as 
\[
x^{(m)}[n] =  x[ (n + m) {\rm mod} N].
\]
Thus, while defined for all integer $m$, there are $N$ unique cyclic shifts.

The \emph{cyclic (periodic) autocorrelation} of a length $N$ sequence $x[n]$ is defined as 
\[
R_{xx}[\tau] = \sum_{n=0}^{N-1}  x^*[n] x^{(\tau)}[n],
\]
for $\tau = 0, 1, \ldots N-1$.    Note that $x^*[n]$ means the complex conjugate of $x[n]$.

The \emph{normalized cyclic autocorrelation} is 
\[
\bar{R}_{xx} [\tau] = \frac{1}{N} R_{xx}[\tau],
\]
which results in $\bar{R}_{xx}[0] = 1$ and $\bar{R}_{xx}[\tau \neq 0] \leq 1$.

The \emph{cyclic (periodic) cross correlation} of two sequences $x[n]$ and $y[n]$, which are each of length $N$, is
\[
R_{xy}[\tau] = \sum_{n=0}^{N-1}  x^*[n] y^{(\tau)}[n],
\]
for $\tau = 0, 1, \ldots N-1$.      For $y[n] = x[n]$ this reduces to the above autocorrelation function.  

Similarly the \emph{normalized cyclic cross correlation} is 
\[
\bar{R}_{xy}[\tau] = \frac{1}{N} R_{xy}[\tau].
\]
Note that it is common in practice to call the single value $\bar{R}_{xy}[0]$ the ``cross correlation'' since this gives the relative correlation between the two sequences without any shifting.   Also, please note that the above definitions are all for discrete time sequences.   In continuous time, the possibility of a partial non-integer time offset becomes pertinent (and in fact likely), and the above definitions would need to be generalized to include continuous time offsets.  We neglect this complication herein.

\subsection{Key Properties}

ZC sequences have several highly desirable properties.

\textbf{Property 1: Constant Amplitude.}  Obviously, from \eqref{eq:defn} since they are unit amplitude complex numbers, all values of $s_q[n]$ have a constant amplitude of 1 and only the phase changes from sample to sample.   This is desirable for several implementation reasons, for example the peak to average power ratio of the sequence is also 1, and also limits the PAPR of the eventual continuous time signal as well.   Since the amplitude need not be computed, only the phase needs to be stored.  

\textbf{Property 2: Zero Cyclic Autocorrelation.}  The cyclic autocorrelation  of a ZC sequence is optimal, in that it is zero for all nonzero shifts of the sequence.   Namely, the unnormalized cyclic autocorrelation function of $s_q[n]$ is $\Nzc \delta[\tau]$, where $\tau \in \mathbb{Z}$ is the cyclic shift.      The normalized cyclic autocorrelation function of $s_q[n]$ is $ \delta[\tau]$.   Again, we caution that in continuous time, with a partial shift (any fraction of a sample time) of the sequence, this property does not hold.    

These first two properties are sometimes combined to be called Constant Amplitude Zero Autocorrelation (CAZAC).   ZC sequences thus are \emph{CAZAC sequences}.

\textbf{Property 3: Fixed Cyclic Cross-Correlation.}   For any two distinct ZC sequences of the same length -- i.e.  the same value of $\Nzc$ but one having  root index $q=q_1$ while the other with root index $q_2 \neq q_1$ --  the normalized cyclic cross correlation is exactly $1/\sqrt{\Nzc}$.    This assumes that $\Nzc$ is prime,  or more generally that $|q_1 - q_2|$ is relatively prime to $\Nzc$, i.e. the only positive integer evenly dividing $\Nzc$ and $|q_1 - q_2|$ is 1.    The unnormalized cyclic cross correlation is$\sqrt{\Nzc}$.   This is in fact the optimal cross correlation for any two sequences with the optimal autocorrelation defined above.   So it is somewhat remarkable that ZC sequences, provided $\Nzc$ is a prime number, can furnish $\Nzc-1$ of such sequences.    Note that if $\Nzc$ is not a prime number, the $|q_1 - q_2|$ constraint reduces the number of ``good'' sequences, which is why prime numbers are generally preferred.

\textbf{Property 4: The DFT or IDFT of a ZC Sequence is a ZC Sequence.}   Since the DFT of a sequence $x[n]$ is a sum of complex exponentials with rotating phase shifts weighted by the sequence $x[n]$, if the sequence $x[n]$ is a ZC sequence, which itself is a rotating sequence of phase shifts, the result is also a ZC sequence.   Similarly, the IDFT of a ZC sequence is also a ZC sequence.    The exact mapping for the IDFT or DFT of a ZC sequence depends on the length $\Nzc$ of the sequence and can be found in \cite{LiHua07}.

The implementation benefit of this property is that a ZC sequence can be generated directly in the frequency domain without actually taking the DFT of the sequence.    This is particularly useful for Orthogonal Frequency Division Multiple Access (OFDMA) or its cousin SC-FDMA.  These are both waveforms that utilize the frequency domain for signaling.  LTE uses OFDMA in the downlink and SC-FDMA in the uplink, while 5G uses OFDMA in the downlink and can use either one in the uplink.       Note that computing the FFT for sequences of prime (or odd) length is quite inefficient, which makes this property especially appealing.

\subsection{ZC-like Sequences that are not of prime length}
\label{sec:ZC-like}

We end this section by emphasizing that these properties listed above only apply to ZC sequences meeting the above stated conditions.  Most commonly, this means the sequences are of length $\Nzc$ wherein $\Nzc$ is a prime number, although there are additional possibilities as discussed under Property 3.   This limitation is not necessarily convenient.  For example in many practical cases the desired spreading factor will not be a prime number.   An arbitrary length \emph{ZC-like sequence} can be created from a prime length one using either cyclic extension or truncation. However, Properties 2 and 3 no longer hold, although the loss may be tolerable.   Properties 1 and 4 do still hold. 

Confusingly, such truncated or extended ZC sequences are often still called ZC sequences by practicing engineers (and their wireless standards), even though they do not satisfy the above theoretical properties and are not actually ZC or CAZAC sequences.   We explore this more below in Example 2.    We will distinguish these imperfect sequences by denoting them as $z_q[n]$, and will refer to them as ``ZC-like''.

To create a \emph{cyclically extended} ZC-like sequence of arbitrary length $\Nsf$, using \eqref{eq:defn}, compute
\[
z_q[n] = s_q[n ~ {\rm mod}~  \Nzc]
\]
for $n = 0, 1, \ldots, \Nsf-1$ and where $\Nzc$ is the largest prime number less than or equal to $\Nsf$.   That is, we simply wrap the necessary number of values of the sequence $s_q[n]$ around to the back of the sequences to ``top it off'' to the desired length. 

To create a \emph{truncated} ZC-like sequence of arbitrary length is also simple.   Just create a ZC sequence of length $\Nzc \geq \Nsf$ and cut off the last $\Nzc - \Nsf$ values of the sequence.   In practice, cyclically extended sequences tend to be preferred to truncated sequences for reasons that are not entirely clear to me.  Perhaps it is because fewer unique samples of the sequence need to be computed or stored.

\section{Examples}

\textbf{Example 1: Length 5 ZC sequences.}   To begin with a simple example, we consider $\Nzc = 5$ and $q=1$.  This gives the ZC sequence 
\begin{eqnarray*}
s_1[0] &=& \exp(0) = 1\\
s_1[1] &=& \exp(-j2\pi/5) \\
s_1[2] &=& \exp(-j6\pi/5) \\
s_1[3] &=& \exp(-j12\pi/5) = \exp(-j2\pi/5) \\
s_1[4] &=& \exp(-j4\pi) = 1 
\end{eqnarray*}
The sequence may seem strange, as the values $1$ and $\exp(-j2\pi/5)$ each appear twice.  But indeed, one can readily verify that this sequence, when multiplied by any shifted and conjugated version of itself and summed, gives 0.   For example, define vectors for the $q=1$ case with shifts 0 and 2 as  
\begin{eqnarray*}
\bs_1^{(0)} &= [1 ~~ \exp(-j2\pi/5) ~~ \exp(-j6\pi/5)~~  \exp(-j2\pi/5) ~~ 1]\\
\bs_1^{(2)} &= [\exp(-j6\pi/5)~~  \exp(-j2\pi/5) ~~ 1 ~~ 1 ~~ \exp(-j2\pi/5)]
\end{eqnarray*}
One can readily compute, with the assistance of Euler's formula, that $(\bs^{(0)})^* \bs^{(2)}  = 0 + 0j$, where $\bs^*$ means the complex conjugate transpose of $\bs$.  It is important to not forget that unlike for the more familiar real sequences, the auto and cross correlation for complex sequences involves taking the complex conjugate of one of the sequences.

Meanwhile, the ZC sequence for $q=4$ can be found to be
\begin{eqnarray*}
s_4[0] &=& \exp(0) = 1\\
s_4[1] &=& \exp(j2\pi/5) \\
s_4[2] &=& \exp(-j4\pi/5) \\
s_4[3] &=& \exp(j2\pi/5) \\
s_4[4] &=& \exp(-j4\pi) = 1 
\end{eqnarray*}
Needless to say, this sequence has the same cyclic autocorrelation property as for $q=1$.   Although not very obviously so, the normalized shifted cross correlation of $s_1[n]$ and $s_4[n]$ can readily be computed to indeed be $1/\sqrt{5} = 0.4472$ for all possible such shifts.   

\textbf{Example 2: Length 12 ZC-like Sequences}.    Because LTE and 5G resource blocks are based upon 12 subcarriers, length 12 ZC-like sequences are commonly used in these standards, particularly for uplink control channels and uplink demodulation reference symbols as described in more detail in Sect. \ref{sec:standards}.    These $\Nsf = 12$ sequences are created via cyclic extension of a $\Nzc = 11$ ZC sequence.  

The unshifted $q=1$ and $q=4$ such length 12 sequences are given below, focusing on the relative phase shifts
\begin{eqnarray}
\bz_1^{(0)} &= \exp\left(\frac{j \pi}{11} \cdot [0 ~~ -2 ~~  -6 ~~ 10 ~~  2 ~~ -8 ~~ 2 ~~ 10 ~~ -6 ~~ -2 ~~ 0 ~~ 0]\right)\\
\bz_4^{(0)} &= \exp\left(\frac{j \pi}{11} \cdot [0 ~~ -8 ~~  -2 ~~ -4 ~~  8 ~~ -10 ~~ 8 ~~ -4 ~~ -2 ~~ -8 ~~ 0 ~~ 0]\right)
\end{eqnarray}

\begin{figure}
\centering
\includegraphics[width=4in]{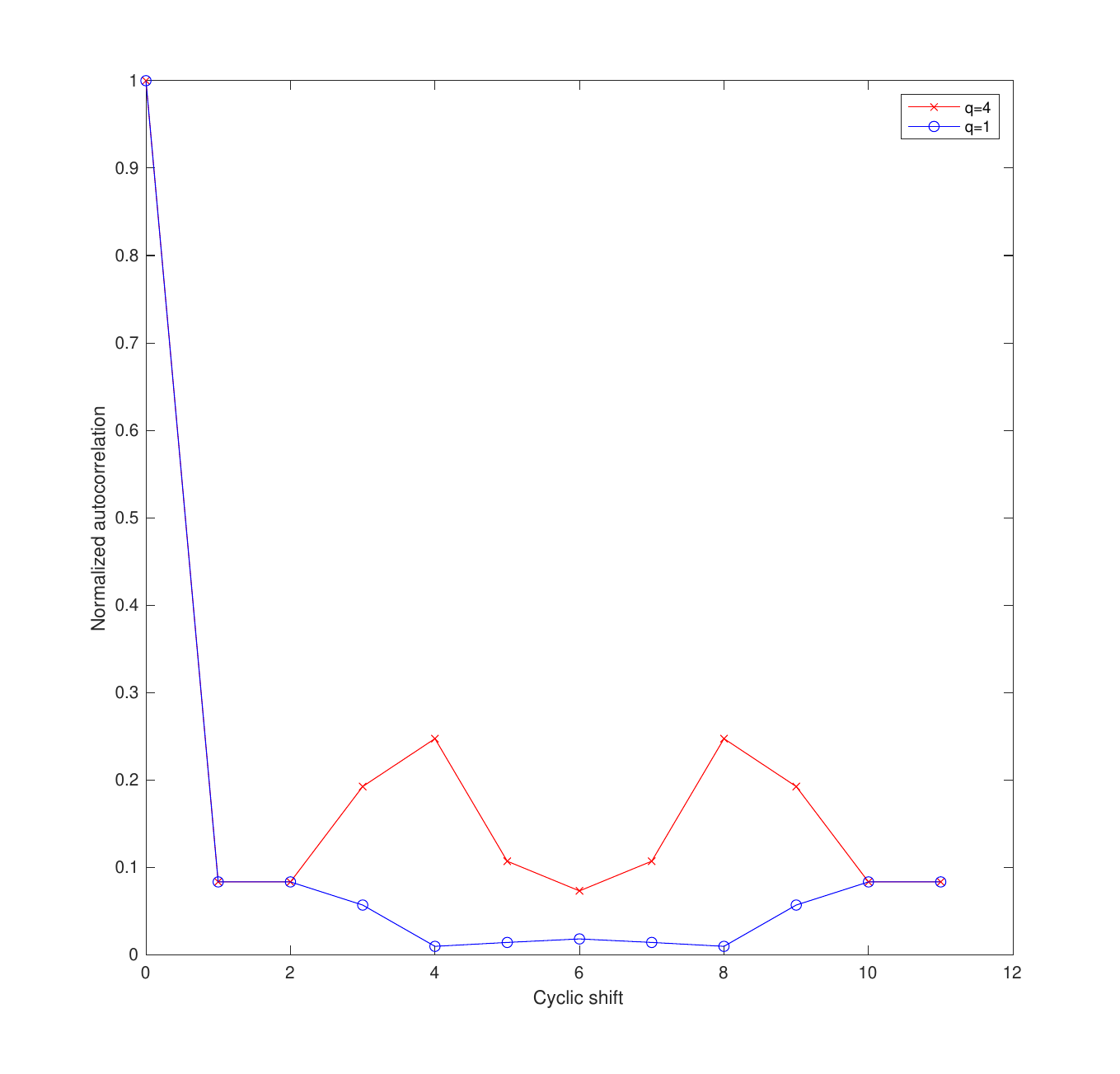}
\caption{Normalized cyclic autocorrelation of the $q=1$ and $q=4$ length 12 ZC sequences.}
\label{fig:autocorr12}
\end{figure}

Both of these sequences,  along with the other (note there are 11 total) different length $12$ cyclically extended ZC sequences, have a normalized autocorrelation that is no longer a delta function, and the autocorrelation of each sequence is different.  Their normalized autocorrelations are plotted in Fig. \ref{fig:autocorr12}, where it can be seen that the autocorrelation for $z_4[n]$ is much worse than for $z_1[n]$.   A common non-zero value for each is $1/\Nsf = 1/12$ (observed at shifts 1, 2, 10, and 11 in the plot), as opposed to the zero autocorrelation for a prime length ZC sequence, although it can be quite a bit higher as seen in the plot.   Recall that in terms of the energy remaining after shifted autocorrelation, these values are squared, which is helpful in terms of maximizing the peak relative to the off-peak terms.

\section{Comparison to Other Common Spread Spectrum Codes}

\textbf{PN sequences}, in particular maximal length PN sequences -- known as m-sequences -- are the most commonly used sequences for spread spectrum.  They are also frequently used for ``scrambling'', which means $\Nsf = 1$, so there is no spreading of the bandwidth and the original digital signal is simply randomized (in a repeatable way).   PN sequences are deterministic $\pm1$ sequences that appear statistically to be nearly identical\footnote{The longer the period of the sequence, the more closely any segment of the sequence resembles a purely random signal} to iid Bernoulli(0.5) sequences, meaning each value of the sequence is 1 or $-1$ with equal probability and independent from all the others.    For two arbitrary shifts or segments of the sequence, correlated over $\Nsf$ values, both the normalized autocorrelation and cross correlations are random variables with variance $1/\sqrt{\Nsf}$.     We recall that the cross correlation \emph{energy} is the square of the normalized cyclic cross correlation, so the energy of interference or noise relative to the desired signal is reduced by approximately $1/\Nzc$ after despreading at the receiver.  Thus the average PN cross correlation is equivalent to the (fixed) cross correlation of ZC sequences of the same spreading factor.    Thus, \emph{on average}, PN sequences have the same cross correlation as ZC sequences, but much higher (worse) autocorrelation.   PN codes also experience fluctuations around that average, although the Law of Large Numbers causes fairly rapid convergence to the mean as the $\Nsf$ increases.   

In summary, PN sequences are widely used and competitive with ZC sequences, but are inferior to ZC sequences in terms of autocorrelation,  and approximately the same in terms of cross correlation.   Their loss stems in large part from the restriction that the PN sequence be real and binary.   When PN codes are applied in complex format -- for example a different PN code is used on each of the I and Q channels -- each functions as an independent PN code: there is no loss nor synergy between them.   In contrast, ZC sequences take advantage of the complex modulation (and in fact require it), to directly encode the phase between the I and Q branches.     Finally, there are several derivate codes of PN codes that combine or modify them in various ways.  The most prominent of these is the Gold Code.

\textbf{Walsh codes} are another commonly used spreading code.  They are orthogonal codes, meaning they have ideal (zero) cross-correlation, providing $\Nsf$ such orthogonal codes for spreading factor $\Nsf$, where $\Nsf$ is necessarily a power of two.   Meanwhile, they have an atrocious (normalized) autocorrelation, not infrequently equal to the maximum value of $1$.  The exact autocorrelation value can be anywhere between 0 and 1 depending on the code and the shift amount.   A similarly poor performance occurs for shifted cross correlations, e.g. when the codes between two users are not perfectly synchronized. This is why Walsh codes are always combined with PN codes when transmitted over a dispersive channel.   A good tutorial on PN and Walsh Codes can be found in the spread spectrum chapter of \cite{GolBook}.

In summary, ZC sequences are complex sequences with unit amplitude and arbitrary phase, as opposed to Walsh and PN codes which are real and binary valued (for our purposes, at $\pm1$).    ZC sequences consist of particular phase shifts of a unit amplitude complex exponential.    In this sense, ZC sequences are more closely related to the otherwise obscure Complementary Code Keying (CCK) modulation used in the first commercially successful WiFi standard, IEEE 802.11b.  

\section{ZC in LTE and 5G}
\label{sec:standards}

ZC sequences are utilized in LTE and 5G NR for many key functions.

\textbf{Initial downlink synchronization}.  This is the first step for a User Equipment (UE) to  begin a connection with a base station (BS).  Synchronization is not only how the UE acquires the BS timing, but also how it receives the key system information necessary for subsequent communication.

In LTE, the Primary Synchronization Sequence (PSS) uses $\Nzc = 63$, and values $q=29, 34, 25$ are possible.  These correspond to a different value of $q$ for each sector in a cellular coverage area, usually.    These are the first signals a UE receives in an LTE frame.

The reasons for choosing an odd but not prime number (such as 61 or 67) for $\Nzc$ are unclear to me, but these are still ZC sequences since 29, 34, and 25 are each relatively prime to 63.  The rationale for choosing those three specific root indices $q$ is based on their superior auto and crosscorrelations under fractional timing and frequency offsets, as demonstrated in \cite{Sesia}.  For example, in the event of a frequency offset of say, 7.5 KHz (half a subcarrier) between either different sectors or due to carrier misalignment at the UE fin a given sector, the desirable theoretical properties of ZC sequences are compromised anyway.   Also, the sequences for $q=29$ and $q=34$ are complex conjugates of each other, so a single correlator can be used to detect both, which allows some desirable complexity reduction at the UE.

In 5G, the PSS is no longer a ZC sequence, but rather is based on PN sequences of length 127.   I believe the reason for the switch is that because the PSS in 5G can be beamformed, and sent in many possible directions.  This is always the case for mmWave, which also has a large bandwidth, but can also be done at lower frequencies.  Therefore, the PSS must be as quickly and easily correlated against as possible, and it is the case that real binary sequences lend to simpler correlators.  

In both standards, the Secondary Synchronization Sequence (SSS) is based on two underlying PN codes.  In 5G it is a Gold Code, while in LTE it is a more complicated interleaving and scrambling procedure based on the two m-sequences.

\textbf{Random access.}  This is how the UE gains access to the network and sends initial information, including possibly a small data payload.   Because the UE has not yet been admitted to the network, it does not have a scheduled time/frequency slot for uplink transmissions.  Thus the uplink random access  channel should be robust to many UEs simultaneously transmitting at slightly different timing offsets: a perfect setting for ZC sequences which have excellent shifted cross and autocorrelation properties.  

Specifically, the Physical Random Access Channel (PRACH) uses $\Nzc = 839$ for the long PRACH preamble and $\Nzc = 139$ for the short PRACH preamble, in both LTE and 5G NR.   These are both prime numbers.

\textbf{Uplink control information.}  The physical uplink control channel (PUCCH) is used by the UEs to convey channel state information and ACK/NAKs, as well as requests to transmit.   All the UEs share the PUCCH, so they periodically use it and also benefit from the use of a spreading code to allow multiple UEs to send control information at the same time and frequency.

In LTE, there are 4 PUCCH formats: 0, 1, 2, and 3.   ZC sequences (length 12, extended from $\Nzc = 11$ sequences as in the above example) are used for all but format 3, which is a supplemental format for use for ultra-high data rates, such as in carrier aggregation, and does not use spreading codes.   

In 5G NR, there are 5 PUCCH formats.    Cyclically extended length 12 ZC sequences are used for formats 0 and 1.
	
\textbf{Uplink reference signals (pilot signals)}.   Uplink reference symbols are transmitted by the UE and utilized by the BS for channel estimation, synchronization, and are necessary for demodulating the UE's data transmission.  In both LTE and 5G, there are two main types of uplink reference symbols:  Sounding Reference Symbols (SRS), which are sent periodically when the UE is not transmitting data,  and Demodulation Reference Symbols (DM-RS) which are embedded with data transmissions to aid with precise channel estimation.

In LTE and 5G NR, length $\Nzc = 31$ ZC sequences are cyclically extended in the frequency domain to achieve length 36 sequences for SRS transmissions.

As far as the DM-RS, whenever SC-FDMA is used, then the DM-RS are length 12 extended ZC-like sequences.    This is always the case in the LTE uplink.   In 5G NR, the uplink can be either OFDMA or SC-FDMA.  When the 5G NR uplink uses OFDMA (which is the default), then a Gold code is used, which aligns with the OFDMA downlink DM-RS in both LTE and 5G.  

\section{Bibliographic Notes}

This primer draws on \cite{Sesia}, which has a summary of Zadoff-Chu sequences and their properties in Sect. 7.2.1, and also discusses their myriad uses in the LTE standard in the relevant sections.     We also gratefully acknowledge the 5G pertinent aspects of ZC sequences discussed in \cite{Dahlman5G}.   

The origin of Zadoff-Chu sequences dates to the late 1950s. My understanding of their origins is as follows.   The ZC sequences themselves were publicly proposed by Chu in 1972 \cite{Chu72}, for any arbitrary length.  However, ZC sequences as known today (i.e. for odd length sequences) were previously proposed much earlier in a secretive patent application \cite{ZadoffPatent} filed in 1957 (and issued in 1963), as noted in \cite{FrankComment} but likely unbeknownst to Chu.   Thus, the dual recognition of both Zadoff and Chu is appropriate, with Zadoff listed first as in \cite{Pop92}, where the name ``Zadoff-Chu" as a descriptor for these sequences appears for the first time, to the best of my knowledge.    



\bibliographystyle{IEEEtran}
\bibliography{Andrews}

\end{document}

%% file: JeffNotation.tex
\def\bs{{\mathbf{s}}}

\def\bz{{\mathbf{z}}}
\def\b0{{\mathbf{0}}}


%% file: Andrews.bib
@article{LiHua07,
	author = {C. {Li} and W. {Huang}},
	journal = {IEEE Transactions on Information Theory},
	number = {11},
	pages = {4221-4224},
	title = {A Constructive Representation for the {Fourier} Dual of the {Zadoff-Chu} Sequences},
	volume = {53},
	year = {2007}}

@article{ZadoffPatent,
	author = {Solomon Zadoff},
	journal = {US Patent 3099796},
	month = {July},
	title = {Phase coded communication system},
	year = {1963}}

@article{Pop92,
	author = {Branislav M. Popovic},
	journal = {IEEE Trans. on Info. Theory},
	month = {July},
	number = {4},
	pages = {1406-1409},
	title = {Generalized Chirp-Like Polyphase Sequences with Optimum Correlation Properties},
	volume = {38},
	year = {1992}}

@article{FrankComment,
	author = {Robert L. Frank},
	journal = {IEEE Trans. on Info. Theory},
	month = {Mar.},
	pages = {244},
	title = {Comments on Polyphase Codes with Good Correlation Properties},
	year = {1973}}

@article{Chu72,
	author = {J. Chu},
	journal = {IEEE Trans. on Info. Theory},
	month = {July},
	pages = {531-32},
	title = {Polyphase codes with good periodic correlation properties},
	year = {1972}}

@book{Dahlman5G,
	author = {Erik Dahlman and Stefan Parkvall and Johan Skold},
	publisher = {Elsevier},
	title = {{5G NR}: the next generation wireless access technology},
	year = {2018}}

@book{Sesia,
	author = {S. Sesia and I. Toufik and M. Baker},
	edition = {2nd},
	publisher = {Wiley},
	title = {{LTE}: The {UMTS} Long Term Evolution},
	year = {2011}}

@book{GolBook,
	author = {A. J. Goldsmith},
	publisher = {Cambridge University Press},
	title = {Wireless Communications},
	year = {2005}}
